\documentclass[11pt]{article}

\usepackage{array,makecell}

\usepackage{amsfonts,amssymb,amsthm,bbm,mathtools}
\usepackage{fullpage}
\usepackage{caption}
\usepackage[colorlinks=true,linkcolor=blue,citecolor=magenta,linktocpage=true]{hyperref}
\usepackage{braket}
\usepackage{titlesec}
\usepackage{cancel}
\usepackage[usenames,dvipsnames]{xcolor}

\usepackage[mathscr]{euscript}
\usepackage{dsfont}
\usepackage{slashed}
\usepackage{currvita}
\usepackage{tocloft}

\usepackage{amsmath}
\usepackage{apptools}

\usepackage{subcaption}
\usepackage{color}
\usepackage{graphicx}
\usepackage{tikz}
\usetikzlibrary{calc}
\usetikzlibrary{decorations.pathmorphing}
\usetikzlibrary{decorations.markings}
\usepackage[numbers,sort,compress]{natbib}

\makeatletter
\renewcommand{\@dotsep}{1000}

\titleformat{\section}{\center\large\bfseries}{\IfAppendix{\appendixname}{} \thesection}{1em}{}

\titleformat*{\subsection}{\flushleft\bfseries}

\definecolor{airforceblue}{rgb}{0.36, 0.54, 0.66}
\definecolor{antiquefuchsia}{rgb}{0.57, 0.36, 0.51}
\definecolor{blush}{rgb}{0.87, 0.36, 0.51}
\definecolor{bondiblue}{rgb}{0.0, 0.58, 0.71}
\definecolor{MyGreen}{rgb}{0.0,0.5,0}
\definecolor{MyDarkRed}{rgb}{0.7,0,0}
\definecolor{MyBlue}{rgb}{0.0,0.0,.5}
\def\be#1\ee{\begin{align}#1\end{align}}
\def\bsub#1\esub{\begin{subequations}#1\end{subequations}}
\def\bg#1\eg{\begin{gather}#1\end{gather}}
\def\ba#1\ea{\begin{eqnarray}#1\end{eqnarray}}

\def\q{\qquad}
\def\f{\frac}
\def\df{\dfrac}

\def\rm#1{\mathrm{#1}}

\def\de{\mathrm{d}}

\newcommand{\R}{{\mathbb R}}
\newcommand{\Z}{{\mathbb Z}}	
\newcommand{\I}{{\mathbb I}}

\newcommand{\cC}{{\mathcal C}}

\newcommand{\cH}{{\mathcal H}}

\newcommand{\cK}{{\mathcal K}}

\newcommand{\cO}{{\mathcal O}}

\newcommand{\cR}{{\mathcal R}}
\newcommand{\cS}{{\mathcal S}}

\newcommand{\ISO}{\mathrm{ISO}}

\renewcommand{\sl}{{\mathfrak{sl}}}

\newcommand{\iso}{{\mathfrak{iso}}}

\numberwithin{equation}{section}

\begin{document}

\title{\Large{\textbf{\sffamily Group quantization of the black hole minisuperspace}}}
\author{\sffamily Francesco Sartini}
\date{\small{\textit{
Univ Lyon, ENS de Lyon, Univ Claude Bernard Lyon 1,\\ CNRS, Laboratoire de Physique, UMR 5672, Lyon, France}}}

\maketitle

\begin{abstract}
The emergence of nontrivial symmetries for black holes minisuperspaces has been recently pointed out.
These Noether symmetries possess non-null charges and hence map physical solutions to different ones.
The symmetry group is isomorphic to the finite-dimensional Poincar\'e group $\ISO(2,1)$, whose irreducible
representations are well known. This structure is used to build a consistent quantum theory of black hole
minisuperspace. This has, among other consequences, the striking consequence of implying a continuous
spectrum for the mass operator. 

Following loop quantum cosmology, we obtain a regularization scheme
compatible with the symmetry structure. It is possible to study the evolution of coherent states following the
classical trajectories in the low curvature regime. We show that this produces an effective metric where the
singularity is replaced by a Killing horizon merging two asymptotically flat regions. The quantum
correction comes from a fundamental discreteness of spacetime, and the uncertainty on the energy of the
system. Remarkably, the effective evolution of semiclassical states is described by an effective
Hamiltonian, related to the original one through a canonical transformation.
\end{abstract}

\thispagestyle{empty}
\newpage
\setcounter{page}{1}

\hrule
\vspace{-0.3cm}
\tableofcontents
\addtocontents{toc}{\protect\setcounter{tocdepth}{2}} 
\vspace{0.5cm}
\hrule

\newpage

\section{Introduction}
Black holes are one of the most fascinating predictions of
general relativity, and the interest in their properties has been
growing since the rise of experimental ability, thanks to
gravitational wave detection and very recent black hole
imagery. But black holes in classical general relativity are
inevitably associated with singularities, signals of the
breakdown of the classical theory. It is widely expected
that approaching the classical singularity, where the curvature becomes Planckian, quantum effects become important.
In a full quantum theory of spacetime, singularities would be
replaced by a unitary evolution through a fuzzy geometry.

Unfortunately, extracting information about the fate of
black hole singularity from a specific quantum theory of
spacetime remains an outstanding challenge. For any
known quantum gravity theory, there is indeed no straightforward way to determine the physical quantum states
representing black holes, even for the simplest case of
spherically symmetric geometries.

It is nonetheless possible to incorporate some features of
the full theories into minisuperspace models with a finite
number of degrees of freedom. Consistent efforts have been
done in this direction, starting from the nonperturbative and
background independent theory of Loop Quantum Gravity
(LQG) and related approaches \cite{Thiemann:2007pyv, Ashtekar:2004eh,rovelli_2004, Perez:2012wv}..

Concerning black hole quantization, the majority of the
works are built on the isometry between black hole interior and Kantowski-Sachs cosmology \cite{Ashtekar:2005qt,Modesto:2005zm,Boehmer:2007ket,Campiglia:2007pb,Chiou:2008eg,Brannlund:2008iw,Chiou:2008nm,Corichi:2015xia,Dadhich:2015ora,Saini:2016vgo,Olmedo:2017lvt,Cortez:2017alh,Ashtekar:2018cay,Ashtekar:2018lag,Bojowald:2018xxu,BenAchour:2018khr,Bodendorfer:2019xbp,Bodendorfer:2019cyv,Bodendorfer:2019nvy,Bodendorfer:2019jay,Ashtekar:2020ckv}, allowing one to
import techniques originally developed in the cosmological
setting. In Loop Quantum Cosmology (LQC), the quantum
effects are claimed to be captured by a phase space
regularization, so-called \textit{polymerization}, that encodes the
fundamental discreteness of spacetime at small scales. In
the first place, the symmetries of spacetime (e.g., homogeneity or spherical symmetry) are imposed classically;
then, quantization is done on a latticelike regularization of
the minisuperspace model, where the essential operators
are holonomies of connection and areas. See e.g., Ref. \cite{Ashtekar:2011ni}
for a review on LQC.

In flat Friedmann-Lemaitre-Robertson-Walker (FLRW)
cosmology, the calculation of expectation values on suitable semiclassical states provides an effective evolution
where the big bang singularity is replaced by a bounce \cite{Ashtekar:2006uz, Ashtekar:2006wn, Bojowald:2001xe}. The effective evolution can be seen as generated by a
regularized classical Hamiltonian, where the canonical
momenta (say $p$) are replaced by the polymerized version
$\sin (\lambda p)/\lambda$, where $\lambda$  is a UV cutoff, typically related to the
Planck length, that could be phase space dependent.

As with any canonical quantization, this procedure suffers
from ambiguities. On top of the usual ordering issues,
polymerization adds another level of ambiguity. To define
the size of the lattice, we should indeed specify a phase space
(in)dependence of the regularization parameter(s). Different
choices of UV regularizations lead to drastically different
dynamics \cite{Corichi:2011pg,Dupuy:2016upu,Corichi:2013usa}. In FLRW models, there is a solid
consensus in favor of the choice know as $\bar \mu$-scheme, even if some ambiguities still remain \cite{Taveras:2008ke,Perez:2005fn,Vandersloot:2005kh,BenAchour:2016ajk}.

Despite the large effort, for the black hole dynamics, no
agreement has been found yet \cite{Ashtekar:2018cay,Ashtekar:2018lag,Bojowald:2018xxu,BenAchour:2018khr,Bodendorfer:2019xbp,Bodendorfer:2019cyv,Bodendorfer:2019nvy,Bodendorfer:2019jay,Ashtekar:2020ckv,BenAchour:2020mgu,Kelly:2020uwj,BenAchour:2020gon,Kelly:2020lec,Husain:2021ojz,Munch:2020czs,Munch:2021oqn}. Moreover,
the majority of the works about black hole dynamics start
directly by the heuristic effective dynamics introducing by
hand the polymer correction. In the absence of fully
controlled quantum dynamics, the question of the equivalence between the effective evolution and the expectation
values of quantum states is rarely addressed.

Nevertheless, it is worth noting that, despite the technical
differences between the various approaches, the effective
models share common features, like the replacement of the
singularity with a transition surface from a black hole to a
white hole interior.

An important question that is raised by the effective
dynamics approach is whether the resulting spacetime
posses covariance. In a more general setting (inhomogeneous), this consistency check is provided by the requirement that the constraints algebra remains anomaly free after
regularization \cite{Tibrewala:2013kba, Bojowald:2015zha}. In the homogeneous minisuperspace
model, this criterion is useless, as we are left only with a
sole scalar constraint, trivially commuting with itself.

In a recent work \cite{Geiller:2020xze}, it has been unraveled that the black
hole interior homogeneous model actually posses a non-trivial and finite-dimensional symmetry algebra, isomorphic to the $\iso(2,1)$ Poincar\'e algebra, that fully encodes
the dynamics on the phase space. This is a generalization of
what happens in flat FLRW cosmology coupled with a
scalar field, in the isotropic case and for the Bianchi I model
\cite{BenAchour:2017qpb,BenAchour:2018jwq,BenAchour:2019ywl,BenAchour:2019ufa,BenAchour:2020njq,BenAchour:2020ewm,BenAchour:2020xif,Achour:2021lqq}, and it has very recently extended to (A)dS Schwarzschild solutions \citep{Achour:2021dtj}.

This invariance has been used as a guiding principle to
build a polymerization scheme in which the symmetry is
protected, replacing the argument of preservation of the
constraint algebra in the full theory. The new symmetry
also opens the door to a group quantization of the model, in
the spirit of what has been developed for the cosmological
model \cite{BenAchour:2018jwq, BenAchour:2019ywl}. The present work aims to exploit the
representation theory of the 2+1 Poincar\'e group to build
a quantum theory of the black hole interior.

The outline of the paper is as follows: I start by reviewing the classical setup for black holes minisuperspace in Sec.	\ref{classic_review}. I recall there the construction of Ref. \cite{Geiller:2020xze}, embedding a massless and zero spin realization of the $\iso(2,1)$ algebra into the mechanical phase space. This algebra corresponds to the evolving version of the Noether charges of the $\ISO(2,1)$ symmetry \cite{Geiller:2020xze}. Arguing that such asymmetry could be used as a guiding principle for quantization, we build a quantum theory of black hole minisuperspace in Sec. \ref{iso_irreps}. Thanks to the knowledge of the energy eigenstates we can impose the dynamics (Sec. \ref{physical space}) on the Hilbert space. This is done in two different ways by respectively strongly and weakly fixing the energy level of the system, and the resulting evolution of the coherent states leads to substantially different effective metrics. The group quantization also gives a very strong prediction about the continuity of the mass spectrum (\ref{mass_spectra}). I end this work with a last section, Sec. \ref{polymerization}, with an analysis of the possible LQC-like deformation of the phase space, such that the symmetry is preserved.

\section{Classical symmetries of black hole minisuperspace}
\label{classic_review}

I start by recalling the notations introduced in the precedent work \cite{Geiller:2020xze}. To study the dynamics of the spherically symmetric homogeneous Kantowski-Sachs cosmology, describing the Schwarzschild interior, we consider the line element
\be
\label{metric_vp}
\de s^2=-N(t)^2\f{V_1(t)}{2V_2(t)}\de t^2+\f{8V_2(t)}{V_1(t)}\de x^2+ L_s^2 V_1(t)\de\Omega^2\,,
\ee
where $x$ runs over the real line and $\de \Omega^2$ is the metric on the unit sphere at constant $x$ and $t$, meaning that the spatial slices have the topology $\R\times S^2$. To get dimensionless fields $V_i$'s we need to introduce a fiducial radius of the two-sphere $L_s$. The system is homogeneous in the sense that the dynamical fields $N$, $V_i$ depend only on time. 

Considering the case of gravity, without matter, and a vanishing cosmological constant, the dynamics is described by Einstein-Hilbert action, integrating the Ricci scalar of \eqref{metric_vp} over a finite slice at constant $t$. For this, we must introduce another fiducial scale $L_0$ in the non-compact $x$ direction. It plays the role of an infrared cutoff that regulates the integration. Because of homogeneity, the finite slice contains information about the whole spacetime, and the action now describes the dynamics of a mechanical system evolving in time, 
\be
\label{EHt}
\cS_\text{EH}^{(t)}[N,V_i]
&=\f{1}{16 \pi L_P^2}\int \de^4 x\, \sqrt{|g|}\, \cR \\
&= \f{L_0}{L_P^2} \int \de t \, \left[N + L_s^2 \f{ V'_1 ( V_2 V'_1 - 2  V_1  V'_2)}{2 N V_1^2 }+\f{\de}{\de t} \left (\f{L_s^2}{2 N V_1} \big(V_1  V_2\big)' \right )\right]
\,\notag\,,
\ee
where $L_P$ is the Planck length and the prime denotes the derivative with respect to coordinate time $t$. The total derivative at the end is exactly compensate by the contribution of the Gibbons Hawking term associated to constant time hypersurfaces. We will drop it in the following, leaving only the first order Lagrangian to describe the classical dynamics of the system. For the following, we will also consider the dimensionless quantity
\be
\kappa = \f{L_0 L_s^2}{L_P^3}\,,
\ee
which encodes the ratio between IR fiducial length and UV scale of the theory, represented by the Planck length. The diffeomorphism invariance of general relativity has been completely fixed, except for the time reparametrization. We face now two possibilities concerning the role of the lapse and the gauge freedom of the time coordinate \cite{Geiller:2020xze}. On the one hand, we can work in coordinate time $t$ and the equation of motion obtained varying the action with respect to the lapse $N$ will correspond the so-called Hamiltonian constraint. This is explicitly given by
\be
\label{classic_constr}
\f{\delta \cS_\text{EH}^{(t)}}{\delta N}=0\q \Leftrightarrow \q 0\approx \cC_\cH = 
-\f{\kappa L_P}{L_s^2}+\f{\kappa L_P}{N^2}\left[\f{V_2{V'_1}^2}{2V_1^2}-\f{V'_1V'_2}{V_1}\right]\,.
\ee 

On the other hand, this is completely equivalent to introducing a \textit{proper time} gauge $\de \tau = N \de t$. With this redefinition, the lapse completely disappears from the action. Moreover, the potential term coming from the intrinsic curvature of the slice becomes a boundary term, and as such can be simply discarded from the action. The latter becomes
\be\label{new_lapse_action}
\cS_0[V_i]
\equiv \kappa L_P
\int \de \tau \,\f{ \dot{V}_1 ( V_2 \dot{V}_1 - 2  V_1  \dot{V}_2)}{2V_1^2 }\,,
\ee
where now the dot represents the derivative with respect to $\tau$. The scalar constraint $\cC_\cH \approx 0$ is now translated into a relationship between the on-shell value of the physical Hamiltonian of $\cS_0$, the fundamental scale and the IR cut-off. This relation is crucial when inserting the on-shell fields into the line element to recover the right solution. In order to better see the equivalence between the two approaches, we perform the canonical analysis of the two actions. In both cases, the conjugate momenta are given by
\be
&\left | \begin{array}{rll}
P_1&=\kappa L_P\df{V_2\dot{V}_1-V_1\dot{V}_2}{V_1^2}&=\kappa L_P \df{V_2 V_1'-V_1 V_2'}{N V_1^2}\,,\\
P_2&=-\kappa L_P \df{\dot{V}_1}{V_1}&=-\kappa L_P \df{V_1'}{N V_1}\,,
\end{array}\right.
\ee
and the Legendre transform gives the form of the Hamiltonian:
\be
\cS_0&=\int \de \tau\,( P_i \dot V_i - H)\,,\q H=-\f{1}{\kappa L_P} \left (P_1 P_2 V_1 +\f{P_2^2 V_2}{2}\right )\,.
\ee
The canonical Poisson structure is $\{V_i,P_j\}=\delta_{ij}$. The dynamics in the two description is described respectively by an Hamiltonian constraint $\cC_\cH$ for the $t$ dependent theory and a true Hamiltonian $H$ for the evolution with respect to the gauge fixed proper time. They satisfy the equality
\be
\label{N_Hamiltonian}
\cC_\cH \equiv N \left (H-\f{\kappa L_P}{L_s^2} \right) = - \f{N}{\kappa L_P} \left(\f{\kappa^2 L_P^2}{L_s^2}+{V}_1{P}_1{P}_2+\f{1}{2}{V}_2{P}_2^2\right) \,.
\ee
For an arbitrary phase space observable $\cO$, we could equivalently describe the dynamics in terms of $t$ or $\tau$, by computing the Poisson brackets of the observable with the respective Hamiltonian density: $\cO'=\{O,\cC_\cH\}$, $\dot \cO=\{\cO, H\}$. The \textit{on-shell} equivalence between the two dynamics is ensured by the vanishing of the scalar constraint, even for lapse choices that are field dependent\footnote{For the two theories to be equivalent and self-consistent, the fiducial scales must be constant}: 
\be
\cO' =\{\cO,\cC_\cH\} =  N\left  \{\cO,H-\f{\kappa L_P}{L_s^2} \right \} + \cancel{\left (H-\f{\kappa L_P}{L_s^2} \right )}\{\cO,N \} \approx N \{\cO, H\} =\f{\de \tau}{\de t}\dot \cO\,. 
\ee

In the following, we will work in proper time and consider the one-dimensional action $\cS_0$, without any lapse or potential term, but we shall remember to relate the latter to the value of the Hamiltonian. Doing so, the equations of motion of the reduced action \eqref{new_lapse_action} are equivalent to the Einstein equations for the metric \eqref{metric_vp} (with $N=1$). A straightforward calculation leads to the classical solutions \cite{Geiller:2020xze}
\bsub
\be
V_1 &= \f{A}{2 L_P^2 \kappa^2} (\tau-\tau_0)^2\,,\\
P_1 &= \f{2 B L_P^2 \kappa^2}{A  (\tau-\tau_0)^2}\,,\\
V_2 &= \f{B}{L_P \kappa} (\tau-\tau_0)-\f{1}{2L_s^2}(\tau-\tau_0)^2\,,\\
P_2 &= -\f{2 L_P \kappa}{\tau-\tau_0}\,,
\ee
\label{classic_traj}
\esub
where the constraint on the value of the Hamiltonian $H=\kappa L_P/L_s^2$ has already been imposed. The other quantities $A,B,\tau_0$ are integration constants. If  we insert these solutions back into the line element \eqref{metric_vp}, and perform the change of coordinates
\be
\tau-\tau_0 = \sqrt{\f{2}{A}} \f{L_P \kappa}{L_s}\, T\,,\q\q x= \f{L_s}{2 L_P \kappa}\sqrt{\f{A}{2}} r\,,
\label{change_classic}
\ee
we find the standard Schwarzschild black hole interior metric, where the mass is given by
\be
\label{classic mass}
M = \f{B\sqrt{A} L_s^3}{\sqrt{2}L_P^2 \kappa^2} =  \f{B\sqrt{A} L_P^4}{\sqrt{2} L_0^2 L_s} \,.
\ee
Notice that the singularity is located at $\tau=\tau_0$, when both the classical solutions for $V_1$ and $V_2$ vanish, while the horizon is at $\tau-\tau_0 = \f{2BL_s^2}{L_P \kappa}$, where only $V_2$ is zero. Although the action $\cS_0$ has been introduced starting from the black hole interior and should be in principle limited to the range where $\tau$ is timelike, the classical solution for the fields $V_i$ are regular on the whole real line, smoothly crossing the singularity at $\tau=\tau_0$.  

In Ref. \cite{Geiller:2020xze} it is shown that all the dynamical information of the metric fields is encapsulated into a phase space structure isomorphic to the $\iso(2,1)$ Poincar\'e algebra. This is provided by the evolving generators associated with the $\ISO(2,1)$ invariance of the mechanical system described by the action \eqref{new_lapse_action}. The main idea behind the construction in Ref. \cite{Geiller:2020xze} is that solving the Hamilton equations is the same as exponentiating the flow of the Hamiltonian vector field, by iteratively computing the Poisson's bracket with the Hamiltonian $\{..\{V_i, H\},..\}, H\}$. One can check if at some point the iteration closes and forms a Lie algebra.  For both the fields, the algorithm stops at the second step
\bsub
\be
C&:= \kappa L_P \{V_2,H\} = -P_1 V_1 - P_2 V_2\,, &\q \{\{V_2,H\},H\}&=-\f{H}{\kappa L_P }\,, \\
D&:= -\kappa L_P \{V_1,H\} = P_2 V_1\,,  &\q A:=(\kappa L_P)^2\{\{V_1,H\},H\} &= \f{V_1 P_2^2}{2}\,,
\ee
\esub
where $A$ is a first integral of motion, commuting with $H$. As the Hamiltonian trivially commutes with itself, for both the fields, the third iteration (i.e the third derivative in proper time $\tau$) vanishes. The generators can be rearranged into the usual $\mathfrak{iso}(2,1)$ basis
\bsub
\be
&J_z= \f{V_2}{2\lambda} - \lambda \kappa L_P H\,,&  
&K_x= \lambda \kappa L_P H  + \f{V_2}{2\lambda}\,,& 
&K_y = C\,,\\
&\Pi_0 = \f{V_1}{2\lambda}+\lambda A\,,&
&\Pi_x= D\,,& 
&\Pi_y= \f{V_1}{2\lambda}-\lambda A\,,
\ee	
\label{poincare_classic}
\esub
where $\lambda$ is a real dimensionless constant. Computing the Poisson brackets between the 6 generators gives the $\iso(2,1)$ algebra
\be
\label{Poincare_alg}
\{J_z,K_i\}& = \epsilon_{ij}K_j\,,&
\{K_x,K_y\}&= -J_z\,,&
 &\cr
\{J_z,\Pi_i\}&=\epsilon_{ij}\Pi_j\,, &
\{K_i,\Pi_0\} & = \Pi_i \,,&  
\{K_i,\Pi_j\} & = \delta_{ij} \Pi_0\,.
\ee
The two Poincar\'e Casimirs, representing respectively the \textit{mass} and \textit{spin} of the corresponding irreducible representation, are
\be
\mathfrak{C}_1= -\Pi_0^2+\Pi_x^2+\Pi_y^2\,, \q\q \mathfrak{C}_2= J_z \Pi_0 +K_x \Pi_y- K_y \Pi_x \,.
\ee
Rewriting the generators in terms of canonical variables we find that the two Casimirs identically vanish $\mathfrak{C}_i = 0$. This condition is necessary to reduce the six-dimensional Lie algebra back to the original four-dimensional phase space. We have already included one of the first integrals of motion ($A$) into the algebra. The other one is related to the Casimir operator of the $\mathfrak{sl}(2,\R)$ sector as
\be
\mathfrak{C}_{\mathfrak{sl}(2,\R)} =-J_z^2+K_x^2+K_y^2:= B^2\,,\q\q B:= V_1 P_1\,,
\ee
that by construction commutes with the Hamiltonian. We shall nonetheless remark that it is not a Casimir operator of the whole algebra and does not select the representation of the algebra, unlike what happens in cosmology \cite{BenAchour:2019ywl, BenAchour:2019ufa} and conformal mechanics \cite{deAlfaro:1976vlx}. 

The intriguing property of this structure is that it can be exponentiated to an actual symmetry of the Lagrangian, generated by the Noether charges corresponding to the initial conditions of the $\iso(2,1)$ generators (see Ref. \cite{Geiller:2020xze} for further development, and Ref. \cite{BenAchour:2019ufa} for the corresponding construction in the isotropic setting). 

It is interesting and useful for what follows to rewrite the (squared) mass in terms of Poincar\'e generators:
\be
M^2= B^2 A \f{L_P^8}{2 L_0^4 L_s^2} := L_M^2(-J_z^2+K_x^2+K_y^2)\f{\Pi_0-\Pi_y}{4\lambda}\,,\q L_M:= \f{L_P^4}{L_0^2 L_s}\,.
\label{mass_Poincare}
\ee
For the sake of simplicity, I introduce $L_M$, which is a constant length indicating how the UV fundamental length and the fiducial scales couple into the definition of the mass observable. We shall remark that, despite the apparent dependence on the fiducial scale, the mass is unchanged by a rescaling of the IR length, because also $A$ and $B$ change under the rescaling \cite{Geiller:2020xze}.

The quantization of the mechanical system representing the black hole interior will be the subject of the next session, but we can already see from here that the vanishing of the Casimir $\mathfrak{C}_i$ implies that we will deal with massless and spin zero representations of the 2+1 Poincar\'e group.

Before moving to the quantum theory, we would like to spend some words about the role of the IR regulator and the presence of boundaries. First of all, we needed to introduce $L_0$ for the action principle of \eqref{EHt} to be well defined. The presence of the second length scale $L_s$ is just a matter of convenience to deal with dimensionless quantities, and allows translating from $L_0$ to $\kappa$. These length scales appear in the energy level for the Hamiltonian, but once we put the solution into the line-element \eqref{metric_vp}, we recover the one-parameter family of black holes with a general mass $M$, as expected from no-hair theorem. 

We have then an apparent clash between the general relativity (GR) point of view (with a one-parameter family of solutions due to no-hair theorem) and the mechanical point of view with a four-dimensional phase space. Furthermore, the symmetries of the mechanical models presented in Ref. \cite{Geiller:2020xze} interplay with the fiducial scales. This is because the symmetry changes the energy value of the solution, and so we must rescale the length in order to restore the constraint \eqref{classic_constr}.  This is actually a quite common situation in gravity, where the boundary seems to carry some physical information, making some gauge redundancy to become physical. Nonetheless, the precise meaning of the physical relevance of the boundary is yet to be determined. For example, some recent work \cite{Achour:2022syr} has pointed out a possible relationship between the conformal properties of the Schwarzschild background and its static perturbation. We will see in the following how the fiducial lengths control the quantum modifications to the classical line element.

\section{ISO(2,1) group quantization}
\label{iso_irreps}
It is a common expectation that the quantum theory of general relativity should provide an anomaly free representation of the constraint algebra. Going into the study of minisuperspaces, such a requirement is meaningless because we are left with the scalar constraint alone, trivially commuting with itself. The unveiling of the conformal symmetry for cosmology and its extension to black holes suggest that for minisuperspaces we can replace the constraint algebra with the new symmetry algebra. In the case considered in this article, this means that the quantum Hilbert space must contain an irreducible representation of Poincar\'e algebra. It is then smart to directly start by exploiting the well known irreducible representation of (the universal cover of) $\ISO(2,1)$. For this, the reader shall refer to an exhaustive discussion in Ref. \cite{Grigore:1993}, of which I recall here some key features in the appendix \ref{ISO_irreps}. We consider the realization of the algebra \eqref{Poincare_alg} as self-adjoint operators, acting on wavefunctions on $\R^2$ in polar coordinates, with the scalar product
\be 
\label{Hilbert_scalar}
\braket{\psi|\chi} =\int_0^\infty \de \rho \int_0^ {2 \pi} \de\phi\, \psi^*(\rho,\phi) \chi(\rho,\phi)\,.
\ee
The $\iso(2,1)$ generators are realized as a one parameter family, with $s\in\R (\rm{mod}	\, 2)$
\bsub
\be
(\widehat{\Pi_0}^{(s)} \psi)(\rho,\phi) =& \rho \,\psi(\rho,\phi)\,,\\
(\widehat{\Pi_x}^{(s)} \psi)(\rho,\phi) =& \rho \sin \phi \,\psi(\rho,\phi)\,,\\
(\widehat{\Pi_y}^{(s)} \psi)(\rho,\phi) =& \rho \cos \phi\, \psi(\rho,\phi)\,,\\
(\widehat{J_z}^{(s)} \psi)(\rho,\phi)  =& \left[ i \f{\partial}{\partial \phi}  -\f{s}{2} \right ]\psi(\rho,\phi)\,,\\
(\widehat{K_x}^{(s)}  \psi)(\rho,\phi)  =& \left[i \rho \left(\sin \phi\f{\partial}{\partial \rho} +\f{\cos \phi}{\rho}\f{\partial}{\partial \phi} \right ) -\f{s}{2}\cos \phi \right] \psi(\rho,\phi)\,,\\
(\widehat{K_y}^{(s)}  \psi)(\rho,\phi) =& \left[i \rho \left(\cos \phi\f{\partial}{\partial \rho} -\f{\sin \phi}{\rho}\f{\partial}{\partial \phi} \right ) +\f{s}{2}\sin \phi \right] \psi(\rho,\phi)\,.
\ee
\label{Poincare_quant}
\esub
By a straightforward calculation, we can verify that they satisfy the quantum version of the algebra \eqref{Poincare_alg}, replacing the Poisson bracket by the commutator between operators.
This realization satisfy the condition that both the Casimirs are zero,
\bsub
\be
\widehat{\mathfrak{C}_1} &:= -\widehat{\Pi_0}^2 +\widehat{\Pi_x}^2+\widehat{\Pi_y}^2 \,,\\
\widehat{\mathfrak{C}_2} &:= \f{1}{2} \left (\widehat{J_z} \widehat{\Pi_0} +\widehat{\Pi_0} \widehat{J_z} +\widehat{K_x} \widehat{\Pi_y} +\widehat{\Pi_y} \widehat{K_x}- \widehat{K_y} \widehat{\Pi_x}- \widehat{\Pi_x} \widehat{K_y} \right  )\,,\\
&\widehat{\mathfrak{C}_2}\, \psi(\rho,\phi) = 0 =\widehat{\mathfrak{C}_1}\, \psi(\rho,\phi)\,.\notag
\ee
\esub
We could explicitly calculate the action of the $\mathfrak{sl}(2,\R)$ Casimir, and see that it does not depend on $s$,
\be
({\widehat{\mathfrak{C}}_{\mathfrak{sl}(2,\R)}}^{(s)}  \psi)(\rho,\phi) =(-\widehat{J_z}^2+\widehat{K_x}^2+\widehat{K_y}^2)\psi (\rho,\phi) = -\rho \left(2\f{\partial}{\partial \rho} +\rho\f{\partial^2}{\partial \rho^2} \right) \psi(\rho,\phi)\,.
\ee
Let us finally remark that we can also define the \textit{square root} of this operator, which corresponds to the classical integral $B$:
\be
(\widehat B  \psi)(\rho,\phi) :=i\left (\rho \f{\partial}{\partial \rho} +\f{1}{2} \right) \psi(\rho,\phi)\,,\q\q \widehat{\mathfrak{C}}_{\mathfrak{sl}(2,\R)}=\widehat{B}^2+\f{1}{4}\,.
\ee
The factor $1/4$ represents a quantum correction to the Casimir with respect to its value in terms of the classical integration constant $B$. 

An important remark should be done at this point. The Hilbert space presented here is unitarily equivalent to the so-called Wheeler-DeWitt (WdW) quantization of our system, where we simply promote the fields $V_i$ to multiplicative operators and $\widehat{P_i} = -i \partial_{V_i}$ (see Appendix \ref{WDW equivalence} for the proof of this statement). Whenever we would like to introduce a quantum phase space inequivalent to Wheeler-DeWitt (via e.g. a polymerization), we expect to brake this unitary equivalence, and at first glance, it could seem impossible to preserve the realization of $\iso(2,1)$ on the polymerized space. We will see in Sec. \ref{polymerization} that the solution is provided by mapping the regularized metric coefficients to operators on the \textit{same} Hilbert space, that are not unitary equivalent to the Wheeler-DeWitt operators. For the time being, we stick to the classical mapping between gravitational reduced phase space and Poincar\'e algebra \eqref{poincare_classic} and further study the Hilbert space and how to impose the dynamics at the quantum level.

\subsection{Hilbert space and energy eigenstates}
First of all we search for eigenstate of the rotation generator $J_z$ and the Casimir $B$. These provide a complete basis of the Hilbert space, and are labelled by a real number $B$ and an integer $m$
\bsub
\be
\widehat{J_z}\ket{B,m}&=m\ket{B,m}\,,\q\q m\in \Z\,,\\
\widehat{B}\ket{B,m}&=B\ket{B,m}\,,\\
\widehat{K_\pm} \ket{B,m}&= \left (m\pm \left (\f 1 2+i B\right )\right )\ket{B,m\pm 1}\,,\\
\widehat{\Pi_0} \ket{B,m}&= \ket{B+i,m}\,,\\
\widehat{\Pi_\pm} \ket{B,m}&= \ket{B+i,m\pm 1}\,,
\ee
\label{j_eigen}\esub
where $K_\pm=K_x\pm i K_y$, and $\Pi_\pm=\Pi_y\mp i \Pi_x$ \footnote{In the last two lines the complex shift of the Casimir must be read as a formal replacement of $B$ into the wavefunction, for example
\be
\braket{\rho,\phi|\widehat \Pi_0|B,m}&= \f{1}{2\pi} \f{1}{\sqrt{\rho}} e^{-i (B+i) \log (\rho)} e^{-\f{1}{2}i (2m+s) \phi}= \f{1}{2\pi} \sqrt{\rho}\, e^{-i B \log (\rho)} e^{-\f{1}{2}i (2m+s) \phi} \notag
\ee
Actually the momenta operator $\Pi_i$ acting on a single eigenstate of the Casimir labelled by $B$ maps it to a combination of eingenstates, exactly like the momentum operator acting on a position eigenstate in standard quantum mechanics. 
}.

The formulas above can be verified by direct computation of the realization \eqref{Poincare_quant} on the normalized wavefunctions
\be
\braket{\rho,\phi|B,m} = \f{1}{2\pi} \f{1}{\sqrt{\rho}} e^{-i B \log (\rho)} e^{-\f{1}{2}i (2m+s) \phi}\,,\q\q
\braket{B', m'|B,m} = \delta_{m,m'} \,\delta(B-B')\,.
\ee
We shall remark that the firsts lines in \eqref{j_eigen} provide a representation of the $\sl(2,\R)$ algebra at fixed Casimir, but the presence of the abelian sector, represented by translations of the Poincar\'e group, allows to move between different values of $B$.

An interesting role is played by the parameter $s$, it is very similar to the one played by the superselection parameter in LQC \cite{Ashtekar:2011ni}. With respect to the scalar product \eqref{Hilbert_scalar}, two states with different $s$ are always orthogonal and belong to different irreducible representations of the Poincar\'e group, exactly as two superselected lattices in LQC with respect to the polymer representation of Weyl algebra \cite{Ashtekar:2002sn, Corichi:2007tf}. Hereafter, without loss of generality, I will set $s=0$.

If we want to impose some dynamics on the Hilbert space we must recall that the classical evolution is generated by the Hamiltonian $H$. At the quantum level, it means that we need to search for eigenstates of $K_x-J_z$,
\be
(\widehat H\,\psi)(\rho,\phi) = \f{1}{2\lambda\kappa L_P}(\widehat{K_x}-\widehat{J_z})\psi(\rho,\phi) = E\,\psi(\rho,\phi)\,.
\ee
Hopefully, the associated differential equation has an analytical solution. Diagonalizing with respect to the Casimir $\mathfrak{C}_{\mathfrak{sl}(2,\R)}$, we obtian again a complete basis for the Hilbert space,
\bsub
\be
\widehat H\ket{B,E}&=E\ket{B,E}\,,\\
\widehat B\ket{B,E}&=B\ket{B,E}\,,\\		
\braket{\rho,\phi|B,E} &= \f{\sqrt{\lambda\kappa L_P}}{2\pi} \f{1}{\sqrt{\rho} \sin(\phi/2)}  e^{-i B \log \left (\rho \sin^2(\phi/2)\right )} e^{-i \left (2\lambda \kappa L_P E \cot\left (\f{\phi}{2}\right ) 
\right )}\,,\\
\braket{B', E'|B,E} &= \delta(E-E') \,\delta(B-B')\,.
\ee
\esub
We shall remark that the spectrum of the Hamiltonian is continuous and unbounded from below, exactly as it happens for cosmology and conformal mechanics \cite{deAlfaro:1976vlx}. This should lead to a catastrophic instability when we consider (multiple) interacting systems, but our formalism is valid only for a single static black hole, without any matter content, and thus the question of stability cannot be addressed here. For instance, it does not make any sense to couple with a thermal bath and to look at the partition function $e^{-\beta H}$.

In order to verify the orthogonality, as well as the completeness, of the basis, it is convenient to perform a change from the $(\rho,\phi)$ polarization to a new set of variables, with respect to which the eigenfunctions look like plane waves (see appendix \ref{ISO_irreps} for the realization of the Poincar\'e algebra on the new variables):
\bg
z:= 2\lambda \kappa L_P \cot \left (\f{\phi}{2}\right )\,,\q\q a:= \log \left (\rho \sin^2\f{\phi}{2}\right )\,,\q\q(z,a)\in \R^2\,,\\
\braket{z,a|B,E} = \f{1}{2\pi} e^{-i B a} e^{-i E z 
}\,.
\label{za_wave}
\eg
I choose to call the second variable $a$ because it is actually related to the $A$ operator:
\be
(\hat A\, \psi)(\rho,\phi) = \f{1}{2\lambda} (\Pi_0 - \Pi_y)\psi (\rho,\phi) = \f{1}{\lambda} \left (\rho\, \sin^2\f{\phi}{2} \right )\psi(\rho,\phi):=\f{1}{\lambda}(\widehat{e^a}\, \psi)(\rho,\phi)
\ee
On the other hand, $z$, being conjugated to the energy, is expected to be related to time. This is indeed what happens once we consider physical states satisfying the quantum dynamics. The imposition of the dynamics on the Hilbert space will be the subject of the next subsection.


\subsection{Physical solution and semiclassical states}
\label{physical space}
In the previous section, we found the eigenvectors that diagonalize both the Hamiltonian ($K_x-J_z$) and one of the classical first integral represented by the $\mathfrak{sl}(2,\R)$ Casimir ($B$). We can exploit this basis to impose the dynamics of the system. We face two different possibilities to do so: we recall that at the classical level we can both impose the constraint $\cC_\cH=H-\kappa L_P/L_s^2=0$ or equivalently see $H$ as a true Hamiltonian generating the time evolution with respect to $\tau$. On the quantum level, the two interpretations (hereafter denoted respectively by \textit{strong} and \textit{weak} constraint) will lead to drastically different semiclassical evolutions. In both cases, we will reconstruct the metric as an emergent quantity, based on the expectation values of the fundamental operators.

\paragraph{Strong constraint}
The most natural way of implementing the strong constraint is to require that the physical states are the ones that satisfy $\cC_\cH {\ket\psi}_\text{phys} =0$, or equivalently:
\be
{\ket \psi}_\text{phys} :=\int\de E\int  \de B\, \delta\left (E-\f{\kappa L_P}{L_s^2}\right ) \,\psi (B) \,\ket{B,E} = \int_\R \de B \,\psi (B) \,\ket{B,\f{\kappa L_P}{L_s^2}}	\,,
\ee 
where the energy scales and their ratio $\kappa$ are given \textit{a priori} and they act as multiples of the identity operator. These physical states are of course not normalized within the original Hilbert space, and we need to introduce a new inner product on the physical space. To this purpose, we make use of \textit{group averaging} (or refined algebraic quantization) \cite{Thiemann:2007pyv,Marolf:1995cn,Marolf:2000iq} and define the projector
\be
\label{group_averag}
\delta(\cC_\cH)=\f{1}{2\pi}\int_{-\infty}^\infty \de x \exp\left ( i\, x\,\cC_\cH\right )\,,	\q\q \delta(\cC_\cH): \cH_\rm{kin}\to \cH_\rm{phys}\,,
\ee
that induces the inner product
\be
{\braket{\chi |\psi}}_\text{phys} = {\braket{\chi|\delta(\cC_\cH)|\psi}}_\rm{kin} = \int \de B\, \chi(B)^* \psi(B)\,.
\ee
Notice that on $H_\rm{phys}$, only the quantum operators $\widehat \cO$ that commute with the constraint are well defined $[\widehat \cO,\widehat{\cC_\cH}]=0$; otherwise, their action will map out of the physical subspace. 
We need then to deparametrize the dynamics with respect to a preferred clock (here $\tau$). More explicitly, this means that the observables we can measure are one-parameter families, e.g.,
\be
\label{Dirac_obs}
\widehat{V_1}(\tau) = \f{\tau^2}{2L_P^2 \kappa^2} \widehat {A}\,,\q\q \widehat{V_2}(\tau) = \widehat {B}\, \f{\tau}{L_P \kappa} -\f{\tau^2}{2L_P \kappa} \hat {H}\,.
\ee
From the expression \eqref{za_wave}, we can infer that $B$ and $a$ are conjugated variables, while $z$ is conjugated to the energy $E$, and it is thus traced out in the group averaging \eqref{group_averag}. The most convenient way of representing the physical space is in terms of functions of the variable $a$ (i.e. Fourier transform of $\psi(B)$), upon which the physical observables in \eqref{Dirac_obs} act as
\be 
\hat A \psi(a) = \f{1}{\lambda} e^a \psi(a)\,,\q\q \hat{B}\psi (a) = i \partial_a \psi(a)\,,\q\q \hat{H} \psi(a) = \f{\kappa L_P}{L_s^2} \psi(a)\,.
\ee

The semiclassical states can be obtained by picking as $\psi(a)$ a Gaussian distribution peaked around some classical values $(a_*,B_*)$,
\be
\psi_*(a) = \,\f{1}{(2\pi \sigma^2)^{1/4}}\,e^{-\f{(a-a_*)^2}{4 \sigma^2}} e^{-i B_* a}\,,\q\q
\left|
\begin{array}{rl}
\braket{\widehat{V_1}(\tau)}&=\dfrac{e^{a_*+ \f{\sigma^2}{2}}}{2\lambda} \,\dfrac{\tau^2}{2L_P^2 \kappa^2} :=\dfrac{A_*}{2}\dfrac{\tau^2}{2L_P^2 \kappa^2}\,,\\
\braket{\widehat{V_2}(\tau)}&=\dfrac{ B_*\tau}{L_P \kappa} -\dfrac{\tau^2}{2L_s^2} \,.\\
\end{array}
\right .
\label{v_strong_evolution}
\ee
This means that the expectation values follow the classical trajectories, up to a constant rescaling of the first integral $A$, due to quantum indetermination. This comes without much surprise, as the classical evolution has already been imposed in the deparametrization of the dynamics, in the definition of the one-parameter family of Dirac observables \eqref{Dirac_obs}.

\paragraph{Weak constraint}
The other possibility to impose the dynamics consists in asking that the constraint is satisfied in a weaker sense: $\braket{ \psi|H|\psi}= \f{\kappa L_P}{L_s^2}$. One could imagine that somehow this would account for some deep fuzziness of the geometry, contributing as an effective stress-energy tensor, that allows some fluctuations around the classical constraint. In the following, we will see how this statement should be correctly interpreted, the uncertainty on the energy level coming explicitly into the game.

The imposition of the weak constraint is easily achieved by Gaussian wavepackets, peaked on some semiclassical values for the pairs of conjugated variables $(B_*, a_*)$ and $(\kappa L_P/L_s^2, z_*)$:
\bsub
\be
\ket {\psi_{*}}&:= \int\de E\int  \de B \,\f{1}{(2\pi \sigma_B \sigma_E)^{1/2}}\,e^{-\f{(B-B_*)^2}{4 \sigma_B^2}} \,e^{-\f{(E-\kappa L_P/L_s^2)^2}{4 \sigma_E^2}} \,e^{i B a_*}e^{i E z_*} \,\ket{B,E}\,,\\
\braket {a,z|\psi_{*}}&=  \,\sqrt{\f{2\sigma_B \sigma_E}{\pi}} \,e^{-\left (a-a_* \right )^2 \sigma_B^2} \, e^{-\left (z- z_*\right )^2 \sigma_E^2}\, e^{i B_*\left (a_*- a\right )} e^{i (z_*- z )\kappa L_P/L_s^2 }\,.
\ee 
\label{semiclass:weak}
\esub
Now, all the operators corresponding to the $\mathfrak{iso}(2,1)$ generators are well defined.  If we want to reconstruct the evolution of the black hole, we simply need to map these generators back to the gravitational phase space:
\be
\widehat{V_1}\, \psi(\rho, \phi) =\lambda (\widehat{\Pi_0}+\widehat{\Pi_y})\psi(\rho,\phi)\,,&\q\q
\widehat{V_2}\, \psi(\rho, \phi) =\lambda (\widehat{J_z}+\widehat{K_x})\,\psi(\rho,\phi)\,.
\ee
A straightforward calculation, using the properties of Gaussian integrals, gives the expectation values
\bsub
\be
\braket{\widehat{V_1}}&=\f{1}{2\lambda L_P^2 \kappa^2} e^{a_*+ \f{1}{8\sigma_B^2}}\,\left (z_*^2+\f{1}{4 \sigma_E^2}\right ):=\f{A_*}{2 L_P^2 \kappa^2}\,\left (z_*^2+\f{1}{4 \sigma_E^2}\right ) \,,\\
\braket{\widehat{V_2}}&= \f{B_*}{L_P \kappa} \, z_* -\f{1}{2 L_s^2} \,\left (z_*^2+\f{1}{4 \sigma_E^2}\right )\,.
\ee
\label{expected_gaussian_weak}
\esub
Comparing with the classical solutions, we identify $\tau= z_*$, and we see that the quantum evolution closely follows a classical trajectory, for large time $\tau\gg1/4\sigma_E^2$, up to a shift in the relationship between the classical constant of motion $A_*$ and its quantum realization $e^{a_*}$, due to the quantum indetermination. On the other hand for small $\tau$, the quantum correction comes into play and it actually prevents $V_1$ from being zero, avoiding the singularity. 

We must also notice that the imposition of the weak constraint coincides with considering $H$ as a true Hamiltonian so that the states must satisfy the Schr\"odinger equation
\be
i \partial_\tau \ket{\psi(\tau)} = H \ket{\psi(\tau)}\,,
\ee
whose solution is indeed given by \eqref{semiclass:weak} replacing $z-z_* \mapsto \tau-\tau_0$. In the following Fig. \ref{fig:effective_gaussian}, there is a comparison between the classical solution and the expectation values of a Gaussian semiclassical state.

\vbox{
\begin{center}
\begin{minipage}{0.4\textwidth}
		\includegraphics[width=\textwidth]{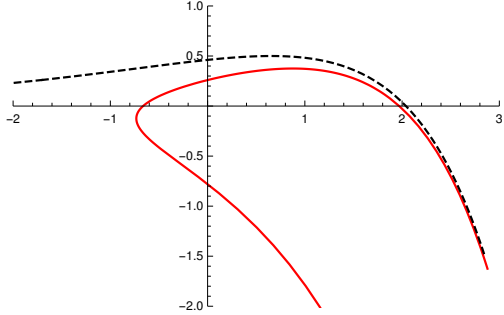}
\end{minipage}
~\hspace{0.5cm}
\begin{minipage}{0.5\textwidth}
\captionof{figure}{\small{Plot of the expectation values of the gaussian state \eqref{expected_gaussian_weak}, with $a_*=1	\,;\lambda=1/\sqrt{2}\,; L_0=1\,; L_s=1\,;\sigma_E=1\,; \sigma_B= 100\,; B_*= 1$, compared with the respective classical trajectory labeled by the first integrals $B=B_*\,,A=A_*$} \label{fig:effective_gaussian}}
\end{minipage}
\end{center}
}
Inserting the expectation values into the line element \eqref{metric_vp}\footnote{We shall remark here that we have actually calculated the expectation values of the fundamental fields $V_i$, and not of the metric coefficients, the two can differ from some $\sigma$ correction. We have considered the effective metric to be:
\be
\de s^2_\rm{eff} = - \f{\braket{V_1}}{2\braket{V_2}}\de \tau^2+\f{8\braket{V_2}}{\braket{V_1}}\de x^2+\braket{V_1}L_s^2\de\Omega^2\,.\notag
\ee
Doing so we have a well defined operator $V_i$, otherwise we should introduce some regularization to deal with the inverse operator.}, with a suitable change of coordinates similar to \eqref{change_classic}\footnote{The change of coordinate is here
\be
\tau= z_*=\sqrt{\f{2}{A_*}}\f{L_P\kappa}{L_s} T\,,\q\q x= \f{L_s}{2 L_P\kappa}\sqrt{\f{A_*}{2}} r \notag\,.
\ee}, we can recast the metric into the form
\be
\de s^2_\rm{eff} = -f(T)^{-1} \de T^2 + f(T) \de r^2 + (T^2 + \Delta) \de \Omega^2\,,\q f(T) =\f{2 M T - T^2 - \Delta}{T^2 + \Delta}\,,
\label{effective_metric}
\ee
where the quantum corrections are encoded in the parameter
\be
\Delta = \f{L_s^2}{8\lambda \sigma_E^2 L_P^2 \kappa^2} e^{a_* + \f{1}{8 \sigma_B^2}} = \f{A_* L_P^4}{8 \sigma_E^2  L_0^2 L_s^2}\,.
\ee
It depends on the scales of the system, but also on the quantum states through the uncertainty on the energy $\sigma_E$ and the classical first integral $A_*$. For small quantum correction ($\Delta/M^2\ll 1$), the region where $\tau$ is timelike is bounded by two horizons for the Killing vector $\partial_r$, in correspondence of the zero of $V_2$. The outer one is close to $T\approx 2M$ and represents the event horizon for the outside of the black hole. The inner one is close to $T\approx 0$. The interior structure resembles closely the Reissner-Nordstr\"om solution of general relativity, bounded by two null horizons. Extending the solution outside the horizons \cite{Ashtekar:2018cay, Achour:2021dtj}, we actually merge two asymptotically flat regions at $T\to\pm\infty$, without any singularity. 

Looking at the exterior region for positive $T$, this effective solution will give a new class of stationary modified black holes, and it would be interesting to further study the correction to standard black hole physics (e.g. Hawking radiation or quasi-normal modes) on such an effective spacetime \cite{Arbey:2021jif, Arbey:2021yke}. Concerning the region behind the inner horizon, it represents a white hole \textit{outside} region, where $\partial_r$ is timelike, and the \textit{effective} mass is negative $-M$. 

Figure \ref{fig:lighcone} represents a schematic diagram for the lightcones structure in the three regions.

\vbox{
\begin{center}
\begin{minipage}{0.4\textwidth}
		\includegraphics[width=\textwidth]{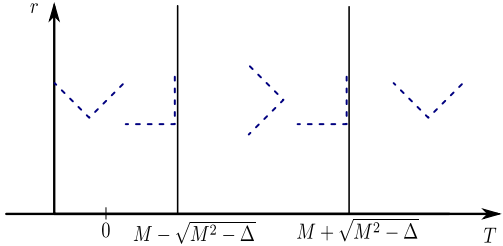}
\end{minipage}
~\hspace{0.5cm}
\begin{minipage}{0.5\textwidth}
\captionof{figure}{\small{Diagram of the future directed lightcones in the effective solution with two horizons, the $T$ coordinates spans the whole real line, the vertical lines correspond to the locations of the horizons} \label{fig:lighcone}}
\end{minipage}
\end{center}
}
We shall nevertheless remark that the locations (and even their existence) of the two horizons depend on the quantum states. Moreover in the extremal limit $\Delta\to M^2$ the two horizons coincide and the quantum correction becomes relevant at a macroscopic scale, meaning that there are large deviations from the classical solution in the low curvature regime, near the horizon. In order to see this, we could also look at the value of the Kretschmann scalar at the transition surface ($T=0$), where the radius of the two-sphere ($V_1$) is minimal. It is given by
\be
\cK_{T=0} = \f{12}{\Delta^2}\,.
\ee
All the corrections to the standard Schwarzschild solution are encoded in the parameter $\Delta$, which in turn depends on the uncertainty on the energy $\sigma_E$ and the classical first integral $A$. The importance of the quantum correction is directly proportional to $\Delta$, so inversely proportional to $\sigma_E$. By calculating the expectation values of the squared operators it is possible to show that the uncertainties on the metric coefficients ($\delta V_i:= \braket{V_i^2}-{\braket{V_i}}^2$) near the minimal radius $(z_*=0)$ also grow inverse proportionally to the dispersion $\sigma_E$. 
\be
\delta V_i\big |_{z_*=0} \xrightarrow[\sigma_E \to 0]{} \cO\left (\f{1}{\sigma_E^4}\right )\,,\q\q
\begin{array}{rl}
\delta V_1\big |_{z_*=0} \xrightarrow[\sigma_E \to \infty]{} \cO\left (\dfrac{1}{\sigma_E^4}\right )\,\\[.4cm]
\delta V_2\big |_{z_*=0} \xrightarrow[\sigma_E \to \infty]{} \cO\left (\dfrac{1}{\sigma_E^2}\right )\,\\
\end{array}\,.
\ee
It is then logical to expect that for a heavily fluctuating metric (small $\sigma_E$), the quantum correction becomes important, and this is indeed what happens. The problem of this model is that for any given dispersion $\sigma_E$, playing with $A$, coupled with the other integral $B$, it is possible to make $\Delta$ as big as desired, without changing $M$. Even for small metric fluctuation (big $\sigma_E$), the deviation from the Schwarzschild solution could be appreciable as close as desired to the horizon, or even cancel the horizon itself (if $\Delta>M^2$). This means that we need to add by hand a first-class constraint on the integral $A$, that fixes its value. We also would like to eliminate the dependence on the fiducial scale of the quantum correction. This uniquely fixes $A_* = L_0^2 L_s^2/L_P^4$ and implies that all the corrections come from the uncertainty on the energy. However, at this point, this constraint might seem a little bit \textit{ad hoc}, and it seems hard to believe that we can infer this kind of constraint from the full theory or its quantization. 

We shall remark that the behaviour of the black hole minisuperspace presents a huge difference with respect to cosmology, where the appearance of quantum correction for the Wheeler-DeWitt quantization has not been observed \cite{Ashtekar:2011ni}. Nonetheless, this dependence on the energy uncertainty disappears for non-squeezed states, where the dispersion on conjugate variables is minimized, by e.g. fixing $\sigma_{E, B} = 1/2$.

We will see in the last section (Sec. \ref{polymerization}) how the results are modified if we introduce a regularization. Before moving to the study of possible regularizations we could exploit again the Poincar\'e structure to discuss the mass spectrum, this is the subject of the next subsection.

\subsection{Mass operator}
\label{mass_spectra}
We recall that at the classical level we have a degeneracy on the definition of the mass, we have indeed two first integral $A$ and $B$, that combine into \eqref{classic mass} to give the only physical quantity that is relevant in the GR framework, the mass, that labels diffeo-inequivalent solutions. But if we look at the quantum theory, we expect that both $B$ and $A$ acquire some fluctuation contributing to the mass. Moreover, the two observables do not commute, and we had to build coherent states to represent semiclassical solutions with a finite spread on both of them. 

This in turn inevitably forces us to work with semiclassical states that are not eigenvectors of the mass operator. Nevertheless, the group quantization provides interesting information about the mass spectrum. For this purpose, we need to further investigate the properties of the mass operator. 

In the first section, we found how we can map the classical observable measuring the squared mass to a combination of Poincar\'e generators \eqref{mass_Poincare}. With the definition of the \textit{square root} of the $\sl(2,\R) $ Casimir, provided at the beginning of this section, we can easily build the self adjoint mass operator
\be
\widehat{M^2} \psi (a,z):=\f{L_M^2}{2} \widehat B \widehat A \widehat B\, \psi(a,z) = -\f{L_M^2}{2 \lambda} \partial_a \left(  e^a \partial_a \psi(a,z) \right )\,
\ee
We shall remark that this operator sees only the $a$ dependence of the wavefunction. Unsurprisingly this means that the mass is a Dirac observable commuting with the Hamiltonian, which in turn implies that it can be measured without any problem on both the weakly and strongly constrained states. For the sake of simplicity, in the following we will consider states on the strongly constrained physical space, tracing out the time-energy dependence of the wavefunction. The conclusions about the spectrum will not be affected by this simplification.

We can explicitly calculate the wavefunctions that diagonalize the mass operator and provide a complete basis for the physical wavefunctions. These are given by the set
\be
\braket{a|M} &:= \psi_M(a)= 2 \f{\sqrt{\lambda M}}{L_M} e^{-a/2}\,\rm{J}_1 \left (\sqrt{8 \lambda} \f{M}{L_M} e^{-a/2}\right )\,,\q \widehat{M^2}\ket{M} = M^2\ket{M}\,. \label{a_mass}
\ee
%
where $M$ is a real positive continuous parameter and $\rm J_1$ is the first order Bessel function of first kind. By virtue of the integral properties of the Bessel functions we can prove the orthogonality and completeness of the basis, with  $y=\sqrt{8\lambda} e^{-a/2}/L_M$, we have indeed
\be
\braket{M'|M}=\int_0^\infty \de y \, \rm J_1(M y) \rm J_1(M' y) \sqrt{MM'} y&= \delta(M-M')\, \q\q
\int_0^\infty \ket{M}\bra{M} =\I\,.
\ee
The existence of the Poincar\'e structure forces the mass to have a continuous spectrum, as it has been pointed out in Ref. \cite{Achour:2021dtj}. This property is in contrast with several other investigations of black hole spectra \cite{Bekenstein:1995ju, Berezin:1998xf} where a discrete spectrum is postulated or obtained \cite{Vaz:1998gm, Louko:1996md, Bojowald:1999ex}. In particular, it means that the black hole could emit particles with any given mass and not only the ones corresponding to the gap between eigenstates.

\section{Singularity and regularization}
\label{polymerization}

In this section, I will discuss how it is possible to define a ``polymer" quantization that preserves the $\ISO(2,1)$ symmetry. For this regularization, the coherent states evolution reproduces the effective metric \eqref{effective_metric} for both the strong and weak constraints.

The main ingredient of Loop Quantum Cosmology is a realization of the Weyl algebra on a non-separable Hilbert space, inequivalent to the standard Schr\"odinger representation. For a given configuration variable (say $q\in \R$), the space is spanned by orthogonal vectors $\ket{q}$, and it contains functions that are non-vanishing only on a countable subset of $\R$. The lack of weak continuity implies that the momentum operator (say $p=-i\partial_q$) is not defined, but only its finite exponential $e^{i \lambda p}$. This leads to the necessity to introduce a regularized Hamiltonian, where the momenta are replaced by (combination of) their exponentiated version. This is usually done by the substitution $p \mapsto \sin (\lambda p)/\lambda$, but other regularizations are possible as well, and the exact form of the effective Hamiltonian has been heavily debated, especially in the context of black hole interior \cite{Ashtekar:2018cay, Ashtekar:2018lag, Bodendorfer:2019xbp, Bodendorfer:2019cyv, Bodendorfer:2019nvy, Bodendorfer:2019jay, Ashtekar:2020ckv}. In any case, the regulator $\lambda$ is claimed to encode the fundamental discreteness of spacetime, relating its value to the Planck length. In the limit where it becomes negligible $\lambda \to 0$, we shall recover the classical evolution. For a given parameter $\lambda$ the Hilbert space is divided into the so-called superselected sector, according to the position eigenstates, the latter taking discrete real values $\epsilon + n \lambda$, with a fixed offset $\epsilon$. The operator $e^{i \lambda p}$ creates a finite shift of step $\lambda$ and lets us move within a given superselected sector. 

The problem with introducing a regularization scheme for the Hamiltonian is that, in general, it spoils the classical Poincar\'e symmetry, unless we extend the regularization to the other observables. A systematic way to ensure that any Poisson structure on a phase space is preserved is to look at the regularization as a canonical transformation \cite{Geiller:2020xze, BenAchour:2019ywl}, where we then replace the new variables $v_i$ at the place of the corresponding \textit{classical} $V_i$ into the line element. This last step is crucial to make the polymerization describe different physics.

We shall remark that a canonical transformation could be implemented by a non-unitary transformation at the quantum level. Nevertheless, we should be capable of rewriting the \textit{polymer} variables in terms of $\iso(2,1)$ generators and calculating their quantum expectation values on both the weakly and strongly constrained wavefunctions. In other words, the difference between the WdW quantization and the polymer one is not seen as the result of Hilbert spaces that are unitary inequivalent, but as the consequence of considering inequivalent operators (not related by a unitary transformation) on the \textit{same} Hilbert space, that in addition carries an irreducible representation of $\ISO(2,1)$.

Nevertheless, we cannot freely choose any transformation, but we want it to satisfy a set of properties:
\begin{itemize}
\item the transformation must be such that the effective metrics is asymptotically equivalent to the Schwarzschild solution,
\item we want that the phase space functions representing the \textit{polymer} coefficients to have a quantum realization with discrete spectra.
\end{itemize}

We already have at our disposal an operator whose eigenvalues are discrete, and with a superselected sector, it is the rotation generator $J_z$; the idea is then to take the regularized metric coefficient to be \cite{Bodendorfer:2020ovt}
\be
v_2 = 2\lambda J_z = V_2 - 2\kappa L_P\lambda^2 H\,.
\label{poly_v2}
\ee
The $\lambda$ parameter must be the same as in the mapping from the original phase space to the $\iso(2,1)$ generator to ensure the right limit $\lambda\to 0$, that maps back to the original phase space. For the superselected sector chosen in the previous sections $(s=0)$, the eigenvalues of $v_2$ are discrete real values $2 n \lambda\,, n \in \Z$.   

Concerning $V_1$, the polymerization is less straightforward. Assuming the transformation \eqref{poly_v2} for $V_2$, we find that a compatible canonical transformation is of the form
\be
\left |
\begin{array}{rl}
v_1&= v_1(\Pi_0, B) = v_1(V_1 + \lambda V_1 P_2^2, V_1 P_1)\,,\\
p_1&= p_1(\Pi_0, B) = p_1(V_1 + \lambda V_1 P_2^2, V_1 P_1)\,,\\
v_2&= 2\lambda J_z = V_2 - 2\lambda^2 \kappa L_P H\,,\\
p_2&= \df{1}{\lambda} \arctan (\lambda P_2)\,,\\
\end{array} \right.
\ee
with two function satisfying $\{v_1, p_1\}=1$. We thus need to find an operator, composed of $\Pi_0$ and $B$ that has a discrete spectrum. Unfortunately, this is not achievable through a linear combination, but we need at least a quadratic operator. The simplest one is given by
\be 
\widehat{v_1^2} := 4 \lambda^2 \hat \Pi_0^2 - \mu^2 \hat B^2\,,
\label{polymer_v1}
\ee
with a real parameter $\mu$. Its discrete eigenvalues (see appendix \ref{v1_eigen} for the technical details) are $ 4\mu^2 n^2\,,n \in \Z$. In this case, we will not have access to the quantum operator measuring $v_1$, but only its square value. From the point of $v_1$, this is similar to what happens in cosmology, where the fundamental discreteness is imposed on the volume, the third power of the scale factor. From the point of view of the scale factor in FLRW cosmology and $v_1$ here, it looks like a so-called $\bar \mu$-scheme. Nevertheless, on the semiclassical level, it is possible to take the square root and implement the canonical transformation: 
\be
\left |
\begin{array}{rl}
v_1&= \sqrt{4 \lambda^2 \Pi_0^2 - \mu^2 B^2} = V_1 \sqrt{(1+P_2 \lambda^2)^2 - \mu^2 P_1^2}\,,\\
p_1&= \f{1}{\mu} \arctan \left (\df{\mu P_1}{\sqrt{(1+P_2 \lambda^2)^2 - \mu^2 P_1^2}}\right )\,,\\
v_2&= V_2 +\lambda^2 P_2 (2 P_1 V_1 + P_2 V_2)\,,\\
p_2&= \df{1}{\lambda} \arctan (\lambda P_2)\,.\\
\end{array}\right.
\ee
For the previous construction to make sense, we need to add a constraint on $\mu$. The minimal value of $\Pi_0$ during the classical evolution is provided by $2\lambda A$. If we want a definite positive square $v_1^2$, we need
\be
\mu  \leq \f{2 \lambda^2 A }{B}\,,
\ee
when the strict inequality holds, we have that $v_1$ is never zero and there is no singularity in the effective metric. On the other hand, if  $\mu B =2 \lambda^2 A$ there is a singularity at $T=0$. If the inequality is not satisfied, the canonical transformation is not defined on the whole trajectory and the singularity occurs before the classical one. The only way to have a canonical transformation that is well defined on the whole phase space is to take the limit $\mu\to 0$, which corresponds to do not polymerise $V_1$. This is somehow analogous to what has been found for the deformation of the constraint algebra for spherically symmetric spacetimes \cite{Arruga:2019kyd}, where only one of the momenta gets polymerized. When $\mu$ goes to zero, we gain back an operator measuring directly the metric coefficient $v_1$, instead of its square value, but now it has continuous eigenvalues:
\be
\mu \to 0\q \Rightarrow\q
\left |
\begin{array}{rl}
v_1&= 2\lambda \Pi_0 = V_1 (1+P_2 \lambda^2)\,,\\
p_1&= \df{P_1}{1+P_2 \lambda^2}\,,\\
v_2&= 2\lambda J_z\,,\\
p_2&= \df{1}{\lambda} \arctan (\lambda P_2)\,.\\
\end{array}\right.
\ee
Inverting the canonical transformation we can express $H$ in terms of the polymerised variables and it gives
\be
\kappa L_P H= -v_1 \f{\tan \mu p_1}{\mu} \f{\sin (2\lambda p_2)}{2\lambda} - v_2 \f{\sin^2(\lambda p_2)}{2\lambda^2} \q\xrightarrow[\mu\to 0]{} \q-v_1 p_1 \f{\sin (2\lambda p_2)}{2\lambda} - v_2 \f{\sin^2(\lambda p_2)}{2\lambda^2}
\label{poly_hamilton}
\ee
The evolution of $v_1$ and $v_2$, generated by the effective Hamiltonian, can be easily solved by inverting the canonical transformation above.

The main advantage of looking at the regularization as a canonical transformation is that the description of both classical and quantum mechanics in terms of Poincar\'e generators is not modified. In other words, the evolution is always generated by $K_x -J_z$, and the dynamical quantum states are the same as in the previous section, according to which kind of constraint we want to impose. We simply need to change the operators whose expectation values we want to replace in the semiclassical metric, I study here the case where $\mu \to 0$, and the effective metric corresponds to  
\be
\de s^2_\rm{eff} = - \f{\braket{\Pi_0}}{2\braket{J_z}}\de \tau^2+\f{8\braket{J_z}}{\braket{\Pi_0}}\de x^2+2\lambda \braket{\Pi_0}L_s^2\de\Omega^2\,.
\label{regular_metric}
\ee
Without much surprise, deparametrizing the dynamics with respect to the time $\tau$, the evolution of $J_z$ and $\Pi_0$ on the physical Hilbert space satisfying the strong constraint follows the respective classical trajectories
\be
\begin{array}{rl}
\braket{\lambda  \widehat{\Pi_0}(\tau)}&=\dfrac{A_*}{2\kappa^2 L_P^2}\,(\tau^2 + 4\kappa^2 L_P^2 \lambda^2) \,,\\
\braket{\lambda  \widehat{J_z}(\tau)}&= \dfrac{B_*}{\kappa L_P}\, \tau -\dfrac{1}{2L_s^2}\,(\tau^2 + 4\kappa^2 L_P^2 \lambda^2)\,.\\
\end{array}
\ee
The expectation values are on the Gaussian physical state, as in \eqref{v_strong_evolution}, and the deparametrization has been done by using the classical solution for $\Pi_0$ and $J_z$ given directly by \eqref{classic_traj}, and then we replace the first integrals $A, B$ by the corresponding quantum operator. This leads to the same effective metric as in \eqref{effective_metric}, where now $\Delta= 2\lambda^2 A_*L_s^2$. 

The interesting thing is that the $\sigma$ correction appearing in the evolution on the weakly constrained states  goes in the same direction. More precisely on the weakly constrained Gaussian wavepackets \eqref{semiclass:weak} we have
\bsub
\be
\braket{\widehat{\lambda  \Pi_0}}&=\f{1}{2\lambda L_P^2 \kappa^2} e^{a_*+ \f{1}{8\sigma_B^2}}\,\left (z_*^2+4 \lambda^2\kappa^2 L_P^2+\f{1}{4 \sigma_E^2}\right ):=\f{A_*}{2\kappa^2 L_P^2}\,\left (z_*^2+ 4 \lambda^2\kappa^2 L_P^2+\f{1}{4 \sigma_E^2}\right ) \,,\\
\braket{\widehat{\lambda J_z}}&= \f{B_*}{L_P \kappa} \, z_* -\f{1}{2L_s^2}\,\left (z_*^2+ 4 L_P^2 \kappa^2 \lambda^2 +\f{1}{4 \sigma_E^2}\right )\,.
\ee
\label{polymer_gaussian_weak}
\esub
The effective structure is again given by \eqref{effective_metric}, but now the quantum correction is encoded in 
\be
\Delta = \left (2\lambda^2 L_s^2+ \f{L_P^4}{8 \sigma_E^2 L_0^2 L_s^2}\right ) A_*\,.
\ee
We shall remark that this does not solve the apparent paradox of quantum correction at a macroscopic scale. If we leave $A$ free, even for metrics with small fluctuation (big $\sigma_E$), the inner horizon can come as close as desired to the external horizon. We still need to add a condition on $A$. 

Taking a closer look at the parameter $\Delta$, we see that it is exactly the sum of the one obtained for the weakly constrained WdW states and the strongly constrained polymer states. It is natural to interpret the two contributions as taking into account respectively the quantum uncertainty on the metric coefficients and the deep discreteness of the spacetime. For wavefunctions that are well localised, the first one is negligible compared to the second one, i.e. $\f{\kappa^2}{\sigma_E^2 L_P^2} \ll \lambda^2 $. In this case, we expect that the quantum corrections are of Planck size, meaning that the first-class constraint would impose 
\be
2 \lambda^2 L_s^2 A \approx L_P^2\q \Rightarrow \q
\Delta \approx \left (L_P^2 + \f{\kappa^2}{16 \sigma_E^2 \lambda^2}\right ) \,.
\label{area_constr}
\ee

This in turn implies that it is impossible to fully get rid of the fiducial scale, entering the game through the central charge $\kappa$. It would be worth studying the role of the cut-off scales, looking at them as running renormalization parameters. The imposition of a first-class constraint relating the polymerization parameter to one of the first integrals is analogous to the construction in Ref. \cite{Ashtekar:2018cay}. Despite \eqref{area_constr} being more reasonable than the one imposed in the WdW setup, the question of if and how we can infer this kind of constraint from the full LQG theory is still unanswered. However, we can have a hint about its origin by remarking the presence of $A$ in the coordinate redefinition \eqref{change_classic}. The relationship \eqref{area_constr} is expected to be related somehow to the introduction of a Planck length ruler on spacetime. The impossibility of completely eliminating the dependence of the effective metric on the fiducial scales points again towards a physical role of the boundary. We would like to stress that similar behaviour has been observed in cosmology \cite{Mele:2021gzx}.

In the previous section, we saw that the evolution of the operators measuring the metric coefficients produces drastically different metrics, depending on whether we allow some energy fluctuation or not. Here the main features of the effective metric are the same in both cases, and moreover they agree with the classical line element corresponding to \eqref{effective_metric}. The evolution of quantum coherent states on the polymer Hilbert space follows the effective evolution described by the corresponding polymer Hamiltonian, and this is stable for non zero energy fluctuation. This feature has been used as a consistency check in favour of the robustness of FLRW polymerization and can be here extended to the black hole interior.

\section{Discussion}
The existence of a hidden symmetry, leaving on top of the time reparametrization, has been revealed for some simple and yet physically relevant minisuperspaces. This is for instance the case of cosmology \cite{BenAchour:2017qpb,BenAchour:2019ywl,BenAchour:2019ufa,BenAchour:2020njq,BenAchour:2020ewm,BenAchour:2020xif}, or black hole interior, \cite{Geiller:2020xze,Achour:2021dtj}. In the quantization of the theory we should deal with the presence of this structure, because of
its ability to encode the classical dynamics. Taking here the conservative approach of preserving the classical structure, has provided a criterion to constrain the quantization. Concretely, this means that any Hilbert space we would like to choose, being it the standard Schr\"odinger picture, or a regularized polymer space, must contain an irreducible representation of the symmetry group we want to preserve. In this work, I focus attention on the black hole interior, but the construction can be generalised to any minisuperspace model that exhibits similar symmetries. The question of the existence of such structures for a general minisuperspace is currently under investigation. We can as well beg the question of whether the Poincar\'e group for black holes is the maximal symmetry group or is part of a larger structure.

In the article, the well known irreducible representations of $\ISO(2,1)$ are used to build a consistent quantum theory, providing an explicit example of observables and their spectra. The most striking consequence of this construction is that we obtain a mass operator with a continuous spectrum. This has important consequences on the emission spectra of black holes and is in contrast to what has been postulated in various works on black hole quantum physics. The existence of this \textit{hidden} symmetry could also explicitly play a role in perturbation theory, providing an interesting interpretation in terms of conserved quantities associated to test fields propagating on the black hole background.

On a more concrete playground, I have started with a quantization equivalent to the standard Schr\"odinger representation of Wheeler-DeWitt gravity, calculating the expectation values of the metric coefficients on some semiclassical states. Classically, only a particular combination of first integrals (namely the mass) is physically relevant. On the other hand, if we allow some fluctuation on the energy, we have seen that the effective metric, emerging as a result of the quantum evolution, strongly depends on both the first integrals and the amplitude of the fluctuations. 

In the last section, I propose a \textit{half-polymerized} regularization, reminiscent of the modification allowed in the context of deformed constraint algebra for spherically symmetric spacetime \cite{Arruga:2019kyd}. The apparent puzzle of introducing a discretization on the configuration space, keeping the invariance under Poincar\'e group is solved here by looking at the regularised variables as a set of operators that satisfy the polymer-Weyl algebra on the same Hilbert space as the usual Schr\"odinger operator, but the two sets are not related by a unitary transformation.

We find out that the quantum corrections come from two terms going in the same direction, summing up into the parameter $\Delta$ that modifies the classical spacetime structure as in \eqref{effective_metric}. The singularity is replaced by a Killing horizon, leading to a white hole region. The two contributions have been interpreted as the effect of a quantum uncertainty on the metric coefficients, and a constant piece proportional to the Planck length, encoding the fundamental discreteness of spacetime. This implies that the light cone structure is the same for both the weakly and strongly constrained states, where the effective structure is achieved by evolving the metric coefficients with respect to the polymerized Hamiltonian \eqref{poly_hamilton}.

Despite the common feature of replacing the singularity with a black-to-white hole transition, the metric presented in this article is different to the one usually considered for the study of properties of regular black holes (see Refs. \citep{Arbey:2021jif, Arbey:2021yke} and references therein), and it would be interesting to see how this affects the phenomenology.

The existence of the \textit{hidden} Poincar\'e structure has been recently extended to the case with a nonvanishing cosmological constant \cite{Achour:2021dtj}, both for de Sitter and anti-de Sitter cases, meaning that the results of this article are easily generalisable in the presence of a cosmological constant. 

Finally, a puzzling role is played by the boundaries. On the one hand, the boundary seems not to play any role in the physics of the system, because of homogeneity the boundary of the spacelike slice trivially carries the same information as the bulk. On the other hand, in order to reduce the action to a mechanical model, we need to introduce such a boundary to regulate the divergent integration of the action. And it is precisely the IR regulator that interplays with the symmetries, being modified when we act on a physical trajectory  \cite{Geiller:2020xze, Achour:2021dtj}. It also plays a crucial role in the quantum theory, by labelling the solution states. How this could be related to renormalization properties, while we consider the IR regulator as a running parameter, is still an open question. Finally, on an effective level, the fiducial scales (and so the boundary) appear explicitly in the modified line element, suggesting a physical role of the boundary for the quantum effects, and pointing towards some holographic properties of gravity.

\paragraph{Acknowledgements}\ \\
I would like to thank Marc Geiller and Etera Livine for the precious discussions and comments, and Jibril Ben Achour for sharing his enthusiasm regarding the properties of hidden symmetries in minisuperspaces.

\appendix 
\section{Irreducible representations of $ISO(2,1)$}
\label{ISO_irreps}
I report here some key insight on the unitary representation of (the universal cover of) the 3-dimensional Poincar\'e group. I followed the notations of Ref. \cite{Grigore:1993}, where the reader can find a more exhaustive discussion on the properties of such representations. They distinguish into three categories according to the sign of the first Casimir $\mathfrak{C}_1$. The latter represents the mass of the particle when we look at $\ISO(2,1)$ as the symmetry group of the 3D Minkowski space. On top of them, there is the trivial representation, where all the elements of the algebra act as zero on the states. As we are interested here in the massless representations $(\mathfrak{C}_1=0)$, I recall here their main features.

Massless representations are labelled by three parameters $\eta=\pm,\, s \in \R$(mod 2), $t \in \R$. The Hilbert space is usually presented in terms of function on $\R^2$, where the two coordinates represent the spatial components of the momenta, and the scalar product is given by
\be 
\braket{\psi|\chi} =\int_{\R^2} \f{\de x\, \de y}{\rho} \psi^*(x,y) \chi(x,y)\,,\q\q \rho = \sqrt{x^2+y^2}\,.
\ee
And the generators of the algebra act like:
\bsub
\be
({\Pi_\mu}^{(\eta,s,t)} \psi)(x,y) =& p^\mu \psi(x,y) \q\q \q\q p^\mu=(\rho,x,y) \,,\\
({J_z}^{(\eta,s,t)} \psi)(x,y)  =& \left[ -i \left( x \f{\partial}{\partial y}-y \f{\partial}{\partial x} \right ) -\f{s}{2} \right ]\psi(x,y)\,,\\
({k_i}^{(\eta,s,t)}  \psi)(x,y)  =& \left[i \rho \f{\partial}{\partial p^i} - s \epsilon_{ij} \f{p^j}{2\rho}- \f{\eta t}{2}\epsilon_{ij} \f{p^j}{{\rho}^2} \right] \psi(x,y)\,.
\ee
\esub
By a straightforward calculation we can verify that they satisfy the quantum version of the algebra \eqref{Poincare_alg}. As expected, the massless representation gives a zero $\mathfrak{C}_1$, while
\be
\hat{\mathfrak{C}}_2 &:= \f{1}{2} \left (J_z \Pi_0 +\Pi_0 J_z +K_x \Pi_y +\Pi_y K_x- K_y \Pi_x- \Pi_x K_y \right  )\,,\\
\hat{\mathfrak{C}}_2 \psi(x,y) &= \f{1}{2}\, \eta t\, \psi(x,y)\,.
\ee
If now we want to realize the phase space presented in the main text of the article, we shall take $t=0$ to have also a zero $\mathfrak{C}_2$. 

In the paper, we have actually considered another realization of the Hilbert space, unitarily equivalent to the one just presented here. It simply consists in expressing the wavefunction in polar coordinates instead of the Cartesian ones $(x,y) =(\rho \sin \phi, \rho \cos \phi)$, this makes the $1/\rho$ factor disappear from the measure. The relation between the two wavefunction expressions is simply given by a change of variable in the argument.
\be 
\psi(x,y) =\psi(\rho \sin \phi, \rho \cos \phi):= \psi_\rm{pol}(\rho,\phi) \,,\quad\braket{\psi|\chi} =\int_0^\infty \de \rho \int_0^ {2 \pi} \de\phi\, \psi_\rm{pol}^*(\rho,\phi) \chi_\rm{pol}(\rho,\phi)\,.
\ee
By a bit of abuse of notation, we will drop the subscript \textit{pol} and we denote with the same symbol the two realizations. The action of the Poincar\'e generators on these functions is given in the main text at \eqref{Poincare_quant}.

The last realization of the Hilbert space that is used in the main text is the $a-z$ polarisation, or its dual Fourier transform. This turns out to be practical to deal with eigenstates of the Hamiltonian $K_x-J_z$. In this case we still do a change of variables, but now we absorb the Jacobian determinant appearing in the measure into the wavefunction
\bg
z= 2 \lambda \cot \left (\f{\phi}{2}\right )\,,\q\q a= \log \left (\rho \sin^2\f{\phi}{2}\right )\,,\q\q(z,a)\in \R^2\\
\psi_{\rm{az}} (a,z):= \f{1}{\sqrt{\lambda}} e^{a/2} \psi\left (\rho(a,z), \phi(z,a)\right ) \quad\braket{\psi|\chi} =\int_{\R^2} \de a \de z \psi^*(a,z) \chi(a,z)\,.\notag
\eg
We will drop again the subscript, because we always represent the same state, even if the measure for the two realizations is not the same. The argument of $\psi$ will implicitly denote the measure we need to pick up. We conclude the discussion by rewriting the action of the generators in this realization
\bsub
\be
({\Pi_0}^{(s)} \psi)(a,z) =& e^a \left (1+\f{z^2}{4\lambda^2}\right ) \,\psi(a,z)\,,\\
({\Pi_x}^{(s)} \psi)(a,z) =& e^a \f{z}{\lambda} \,\psi(a,z)\,,\\
({\Pi_y}^{(s)} \psi)(a,z) =& e^a \left (\f{z^2}{4\lambda^2}-1\right ) \, \psi(a,z)\,,\\
({J_z}^{(s)} \psi)(a,z)  =& \left[ \f{i}{2 \lambda} \left (z \partial_a+ \f{4 \lambda^2+z^2}{2}\partial_z \right ) -\f{s}{2} - \f{i}{4 \lambda} z \right ]\psi(a,z)\,,\\
({K_x}^{(s)}  \psi)(a,z)  =& \left[ \f{i}{2 \lambda} \left (z \partial_a+ \f{4 \lambda^2-z^2}{2}\partial_z \right ) -\f{s}{2}\f{z^2-4 \lambda^2}{z^2+4 \lambda^2} - \f{i}{4 \lambda} z \right ]\psi(a,z)\,,\\
({K_y}^{(s)}  \psi)(a,z) =& \left[ i \left (- \partial_a+ z\partial_z \right ) +s \f{4 \lambda z}{z^2+4 \lambda^2} + \f{i}{2} \right ]\psi(a,z)\,.
\ee
\esub
And we see that $B$ is simply conjugated to $a$, while $H$ has en extra term depending on $s$ and $\lambda$, that vanishes for the \textit{superselected} sector $s=0$
\be
\hat B \psi(a,z) = i \partial_a \psi(a,z)\,.\q\q
\hat H \psi(a,z) = i \partial_z \psi(a,z)+ \f{2 s \lambda}{4 \lambda^2+z^2} \psi(a,z)\,.
\ee

\section{Wheeler-DeWitt quantization}
\label{WDW equivalence}
In this appendix, we will see how the Poincar\'e irreducible representations are equivalent to the so-called Wheeler-DeWitt quantization, by the latter I mean the natural quantization scheme that consists in promoting the configuration variables $V_i$ to multiplicative operators and the momenta to derivatives $\hat P_i=-i \partial_{V_i}$. The Hilbert space is given by square normalizable wavefunctions $L^2(\R^2,\de V_1 \de V_2)$.

This corresponds to the \textit{position} polarization of the wavefunctions, that is known to be equivalent to the \textit{momenta} polarization up to a Fourier transform. The key idea is to perform the Fourier transform only on the variable $V_2$, meaning that the multiplicative operators are now $V_1$ and $P_2$, and
\be
\begin{array}{rl}
\widehat{V_1} \psi(V_1,P_2)&= V_1 \psi(V_1,P_2)\,,\\
\widehat{P_2} \psi(V_1,P_2)&= P_2 \psi(V_1,P_2)\,,\\
\end{array} 
\q 
\begin{array}{rl}
\widehat{P_1} \psi(V_1,P_2)&= -i \partial_{V_1} \psi(V_1,P_2)\,,\\
\widehat{V_2} \psi(V_1,P_2)&= i \partial_{P_2} \psi(V_1,P_2)\,,\\
\end{array} 
\ee
With these fundamental operators at hand, we can build the observables corresponding to the $\iso(2,1)$ generators on the classical phase space. We simply need to consider the corresponding combination of fundamental operators and make it self adjoint, for example, if we take $K_y=-P_i V_i$, the corresponding hermitian operator is given by:
\be
(\widehat{K_y}  \psi)(V_1,P_2) =& -\f{1}{2}(\widehat{V_1} \widehat{P_1} +\widehat{P_1} \widehat{V_1} + \widehat{V_2} \widehat{P_2} + \widehat{P_2} \widehat{V_2} )\psi(V_1,P_2)\,.
\ee
The whole algebra is then:
\bsub
\be
(\widehat{\Pi_0} \psi)(V_1,P_2) =& \f{V_1(1+P_2^2\lambda^2)}{2\lambda}\,\psi(V_1,P_2)\,,\\
(\widehat{\Pi_x} \psi)(V_1,P_2) =& V_1 P_2\,\psi(V_1,P_2)\,,\\
(\widehat{\Pi_y} \psi)(V_1,P_2) =& \f{V_1(1-P_2^2\lambda^2)}{2\lambda}\,\psi(V_1,P_2)\,,\\
(\widehat{J_z} \psi)(V_1,P_2)  =& i \left[\f{1+P_2^2 \lambda^2}{2\lambda} \partial_{P_2} - P_2 V_1  \right ]\psi(V_1,P_2)\,,\\
(\widehat{K_x}  \psi)(V_1,P_2)  =&i \left[\f{1-P_2^2 \lambda^2}{2\lambda} \partial_{P_2} + P_2 V_1  \right ]\psi(V_1,P_2)\,,\\
(\widehat{K_y}  \psi)(V_1,P_2) =&-i \left[ P_2 \partial_{P_2} - V_1\partial_{V_1} \right ]\psi(V_1,P_2)\,.
\ee
\esub 
This have already the form of the operator in the previous appendix, where the \textit{translation} operator are multiplicative, and the \textit{rotation} and \textit{boosts} contains some derivatives. The unitary equivalence is proven by considering the following change of coordinate in the wavefunction:
\be
\begin{array}{rll}
V_1&= \lambda y +\lambda \sqrt{x^2 + y^2}&=\lambda \rho (1+ \cos \phi)\,,\\
P_2&= -\f{y - \sqrt{x^2+y^2}}{x\lambda}&=\f{1}{\lambda}\tan\f{\phi}{2}\,.\\
\end{array}
\ee
Absorbing the Jacobian of the transformation into the measure of the new realization, we recover the realization presented in the main text and in the appendix \ref{ISO_irreps}, meaning that the two are unitarily equivalent.
\section{$\widehat{v_1}$ eigenstates}
\label{v1_eigen}
In Sec. \ref{polymerization}, I claim that the simplest combination of operators $B$ and $\Pi_0$ that have a discrete spectra is the quadratic functional \eqref{polymer_v1}. In order to see this explicitly we shall in a first place look at the linear combination
\be 
\widehat{v_1} := 2 \lambda \widehat{\Pi_0} + \mu \widehat{B}\,.
\ee
By acting on a wavefunction in the polar representation, we can analytically solve the eigenvalue problem and find the eigenvectors
\be
\braket{\rho,\phi|v_1,m} = \f{1}{2\pi \sqrt{\mu \rho}}e^{-i\left (2\f{\rho\lambda}{\mu}-\f{v_1}{\delta} \log \rho\right )} e^{-i m \phi}\,,\q\q \widehat{v_1}\ket{v_1,m} =v_1 \ket{v_1,m}\,.
\ee
that provide an orthonormal basis for a continuous spectrum $v_1 \in \R$: 
\be
\braket{v_1',n|v_1,m} =\delta_{n,m}\delta(v_1-v_1')\,,
\ee
On the other hand the quadratic combination  \eqref{polymer_v1} considered in the main text has eigenvectors
\be
\braket{\rho,\phi|v_1^2,m} = \sqrt{\f{v_1}{2\lambda\rho}}\,J_{\f{v_1}{\mu}}\left (\f{2 \lambda \rho}{\mu}\right )\, \f{1}{\sqrt{2\pi}}e^{-i m \phi}\,,\q\q \widehat{v_1^2}\ket{v_1^2,m} = v_1^2 \ket{v_1^2,m}\,,
\ee
where $J_n$ are the first kind Bessel fucntions. The eigenstates are normalized for a discrete spectrum $\f{v_1}{2\mu} \in Z$. This is easily shown by using the integral property of the Bessel functions
\be
\int_0^\infty \f{\de y}{y}J_{2n}(y)J_{2m}(y) = \f{1}{2\pi}\f{\sin\left (\pi(n-m)\right)}{n^2-m^2} = \f{1}{2 n}\delta_{n,m}\q \Leftrightarrow\q n,m \in \Z\,.
\ee
\bibliographystyle{Biblio}
\bibliography{Biblio.bib}

\end{document}